\newcommand{\be}{\begin{equation}}
\newcommand{\ee}{\end{equation}}
\newcommand{\ba}{\begin{eqnarray}}
\newcommand{\ea}{\end{eqnarray}}

\newcommand{\Tr}{\hbox{Tr}}
\documentclass[aps,prd]{article}
\usepackage{amsmath}
\usepackage{amssymb}
\begin{document}
\title{The $\star$-value Equation and Wigner Distributions
in Noncommutative Heisenberg algebras\footnote{ Dedicated to Mike Ryan on his sixtieth birthday,
who as a scientist always understood that
it is nice to be good, but that it is better to be nice.} }
\author{Marcos Rosenbaum, and J. David Vergara \\
{\small\it Instituto de Ciencias Nucleares, UNAM, A. Postal
70-543, M\'exico D.F.}}
\date{}
\maketitle

\begin{abstract}
We consider the quantum mechanical equivalence of the Seiberg-Witten map in
the context of the Weyl-Wigner-Groenewold-Moyal phase-space
formalism in order to construct a quantum mechanics over
noncommutative Heisenberg algebras. The formalism is then applied to the
exactly soluble Landau and harmonic oscillator problems in the 2-dimensional
noncommutative phase-space plane, in order to derive their correct energy spectra
and corresponding Wigner distributions. We compare our results with others that have
previously appeared in the literature.
\end{abstract}

\section{Introduction}

There is a fairly deep understanding in theoretical physics
on the microscopic structure of matter, but very little is
known concerning the microscopic structure of the space-time.
We know, for instance, that to distances of the order of $10^{-17} m.$
the space-time is a continuum but we do not know what happens to
distances arbitrarily smaller than that. So, one of the most important open
problems in theoretical physics is to understand the microscopic
structure of the space-time, {\it i.e.} how to build a quantum theory of
gravity.

By means of a simple heuristic argument, based on Heisenberg's
Uncertainty Principle, the Einstein Equivalence Principle and the
Schwarzschild metric, it is easy to show that the Planck length
seems to be a lower limit to the possible precision of measurement
of position, and that shorter distances do not appear to have any
operational meaning. It would then appear reasonable the need to
extend the phase-space noncommutativity of quantum mechanics to a
noncommutativity of space-time in order to quantize gravity.
Furthermore, under these premises the very concept of manifold as
an underlying mathematical structure in the construction of
unified physical theories, applicable to distances of the order of
the Planck length, becomes questionable and some people have been
convinced that a new paradigm of geometric space is needed that
would allow us to incorporate into our theoretical formalisms
completely different small-scale structures from those to which we
are usually accustomed. Among physicists some options for this
paradigm are embodied in topological quantum field theory,
dynamical triangulations, string theory (and efforts in this
context to develop a nonperturbative formulation that could allow
us to reach Planck scale physics) and loop quantum gravity. See
{\it e.g.} \cite{rovelli} for a collection of these different directions of research.\\

Among mathematicians mainly one such outstanding paradigm is the
noncommutative geometry invented by Connes, which considers a new
calculus, the so called spectral calculus, based on operators in
Hilbert space and the use of the tools of spectral analysis
\cite{connes}. This geometry has among its features that it
includes ordinary Riemannian space; discrete spaces are treated on
the same footing as the continuum, thus allowing for a mixture of
the two; it allows for the possibility of noncommuting
coordinates; and even though quite different from the geometry
arising
in string theory, it is not incompatible with it.\\

 Although none of the above mentioned apparently conceptually different approaches
and their variants are anywhere near a final theory of grand unification, and probably
no single one of this directions will succeed in producing it, there appears to be emerging
a common denominator of noncommutativity in some of their ingredients which points to
the fact that when considering the problem of coordinates below the Planck length, there is
no good reason to presume that the texture of space-time will still have a 4-dimensional continuum.\\
Further evidence along this line of thought has been provided by recent developments
in string theory where noncommutative geometry appears in the low energy effective
theory of brane configurations and in the matrix model of M-theory. It has also been shown
recently that in noncommutative field theories the Seiberg-Witten map can be interpreted as a field dependent gravitational background \cite{rivelles}. In fact, it is not difficult
to show that that a similar interpretation can be carried out even at the level of quantum mechanics on noncommutative phase-space.\\
These recent results, as well
as others ({\it c.f.} examples of noncommutative geometry in field theory listed in \cite{alvarez}),
have generated a considerable interest to
understand the role played by noncommutative geometry in different theoretical sectors of physics.

In quantum field theory noncommutativity can be formulated mathematically in two different ways:\\
 1) By means of the  $\star$-product on the space of $c$-functions
\be
 f\star g = \exp (\frac{i}{2} \theta _{ij}\partial_{x_i}\partial_{y_j}) f(x)g(y)|_{x=y},\label{weyl-moyal}
\ee
 or\\
2) By defining the field theory on an operator space that is intrinsically noncommutative.
Although formally well defined, the operator approach is hard to implement in explicit
calculations. Hence the analysis of the noncommutative effects is usually performed by expanding the
 $\star$-product perturbatively.\\

Moreover, since single particle quantum mechanics can be seen, in
the free field or weak coupling limit, as a mini-superspace sector
of quantum field theory where most degrees of freedom have been
frozen   ({\it i.e,} as a one-particle sector of field theory),
the above mentioned results from field theory as well as others
suggest
 that a more detailed study of exactly solvable models in noncommutative
quantum mechanics will be helpful both for the understanding of the effects of
noncommutativity in field theory,
as well as of its possible phenomenological consequences in space.\\
>From the intrinsically noncommutative operator point of view, the
development of a formulation for noncommutative quantum mechanics
requires first a specification of a representation for the
phase-space algebra, second a specification of the Hamiltonian
which governs the time evolution of the system and last a
specification of the Hilbert space on which these operators and
the other observables of the theory act. Regarding the choice of a
representation for the intrinsic Heisenberg noncommutative
phase-space algebra, several works that have appeared lately in
the literature have suggested using a quantum mechanical
equivalent to the Seiberg-Witten map \cite{seiberg}, whereby the
noncommutative Heisenberg algebra is mapped into a commutative one
\cite{nair}, \cite{sochichiu}, \cite{smailagic}, \cite{li}. Since
in all generality this map admits many possible realizations, one
could have in principle also many possible resulting
self-consistent quantum mechanics of which the proper one could
only be discerned by
experiment.\\
As for the choice of the Hilbert space, however, a reasonable
assumption is that it can be taken to be the same as
that for the corresponding commutative system, for any of the
realizations of the noncommutative Heisenberg algebra in terms of
the position and momentum operators
for the commutative one  \cite{chaichian}. \\

The purpose of this paper is to show that a noncommutative quantum
mechanics based on the Weyl-Wigner-Groenewold-Moyal
 formalism, extended to noncommutative phase-space by means of the
 quantum mechanical equivalent of the
Seiberg-Witten map, can provide an interesting frame for further
investigating the above mentioned approaches. In particular, we
analyze the so called Weyl-Moyal correspondence procedure as
symbolized by (\ref{weyl-moyal}), when applying it to two exactly
solvable models: the Landau problem and the harmonic oscillator in
both noncommutative configuration and phase-space. We argue that
this procedure leads to the correct quantum mechanics for the case
of Heisenberg algebras where noncommutativity is restricted to
configuration space and then only when the $c$-Weyl equivalent to
the quantum observables is the same as the ordinary function that
would be obtained by replacing the operators of the commutative
Heisenberg algebra by their corresponding canonical dynamical
variables. In addition, we also show through these examples what
we consider is the correct procedure for applying the
$\star$-value equation (see equation (\ref{starv}) below) to
 the case of non-commutative spaces and for the derivation of
 the Wigner distribution function in this case.\\

In order to make our presentation as selfcontained as possible,
we begin our discussion in Sec.2  with a brief review of
the Weyl-Wigner-Groenewold-Moyal formalism for ordinary quantum
mechanics. We then turn to show how this formalism
can be extended to noncommutative Heisenberg algebras by
resorting to what could be considered a quantum mechanical equivalent
of the Seiberg-Witten map, which we discuss there. In Secs. 3 and 4 we
apply the formalism to calculate the energy spectrum and
Wigner functions for the Landau and harmonic oscillator problems
in noncommutative phase-space as a basis for a
 comparison with the results derived by an application of the
 Weyl-Moyal correspondence and for the analysis of
the particular circumstances when both procedures are equivalent.
We conclude the paper in Sec.5 with some general remarks
on this issues and with suggestions for further work.

\section{Weyl functions and Wigner distributions in commutative and noncommutative
phase spaces}
Let

\be [Q_i, Q_j]  =  0, \nonumber \ee \be [\Pi_i, \Pi_j]  =
0,\nonumber \ee \be [Q_i, \Pi_j]  = i \hbar \delta_{ij},
\label{heisen} \ee be the commutative Heisenberg algebra of
ordinary quantum mechanics. Making use of the
Baker-Campbell-Hausdorff (BCH) theorem one can readily show that
the set of operators $(2\pi\hbar)^{-\frac{d}{2}}
\exp[\frac{i}{\hbar}({\bf x}\cdot {\boldsymbol \Pi} + {\bf y}\cdot {\bf Q})]$ satisfy
the orthonormality condition \be (2\pi\hbar)^{-d} \Tr
\{\exp[\frac{i}{\hbar}({\bf x}-{\bf x}^{\prime})\cdot {\boldsymbol
\Pi} + ({\bf y} -{\bf y}^{\prime}) \cdot {\bf Q})]\} =\delta ({\bf
x}-{\bf x}^{\prime}) \delta ({\bf y}-{\bf
y}^{\prime}),\label{orto} \ee where ${\bf x}$, ${\bf y}$ are
$c$-vectors and $d$ is the dimension of the configuration space.
Thus they form a complete set and any quantum operator $A({\boldsymbol \Pi}, {\bf Q},
t)$ can be written as \be A({\boldsymbol \Pi}, {\bf Q}, t)= \int
\int d{\bf x} \ d{\bf y} \alpha({\bf x}, {\bf y}, t)
\exp[\frac{i}{\hbar}({\bf x}\cdot {\boldsymbol \Pi} + {\bf y}\cdot
{\bf Q})],\label{op} \ee where, by (\ref{orto}), the $c$-function
$\alpha({\bf x}, {\bf y}, t)$ is determined by \be \alpha({\bf x},
{\bf y}, t)=(2\pi\hbar)^{-d} \Tr \{ A({\boldsymbol \Pi}, {\bf Q},
t) \exp[-\frac{i}{\hbar}({\bf x}\cdot {\boldsymbol \Pi}
 + {\bf y}\cdot {\bf Q})] \}.\label{alpha}
\ee
Define now the Weyl function corresponding to the quantum operator
$A({\boldsymbol \Pi}, {\bf Q}, t)$ by
\be
A^{W} ({\bf p},{\bf q}, t)= \int \int d{\bf x} \ d{\bf y} \alpha({\bf x},
{\bf y}, t) \exp[\frac{i}{\hbar}({\bf x}\cdot {\bf p} + {\bf y}\cdot {\bf q})],
\label{weyl}
\ee
and consider the expectation value of the product of two quantum operators
relative to the pure state $|\Psi\rangle$. The Weyl-Wigner-Groenewold-Moyal
phase space formalism then shows that
\be
\langle \Psi |A_{1} ({\boldsymbol \Pi}, {\bf Q}, t) A_{2}
({\boldsymbol \Pi}, {\bf Q}, t) | \Psi \rangle = \int \int d{\bf p} \
d{\bf q} \rho_{W} ({\bf p},{\bf q}, t)
 A^{W}_{1} ({\bf p}, {\bf q}, t) \star  A^{W}_{2} ({\bf p}, {\bf q}, t), \label{wigner}
\ee where \be \rho_{W} ({\bf p},{\bf q}, t) = (2\pi\hbar)^{-d}
\int d{\bf z} \exp[\frac{i}{\hbar} {\bf z}\cdot{\bf p}]\langle
{\bf q} -\frac{{\bf z}}{2} |{\boldsymbol \rho} |{\bf q} + \frac{{\bf
z}}{2}\rangle, \label{dist} \ee is the Wigner quasi-probability
distribution function, ${\boldsymbol \rho} =|\Psi\rangle\langle \Psi|$ is the
von Neumann density matrix for a pure quantum state, and \be \star
= \exp [ \frac{i\hbar}{2} \Lambda ] := \exp \left[
\frac{i\hbar}{2} ( {\overleftarrow{\nabla_{{\bf q}}}}\cdot
{\overrightarrow{\nabla_{{\bf p}}}} - {\overleftarrow{\nabla_{{\bf
p}}}}\cdot {\overrightarrow{\nabla_{{\bf q}}}} ) \right],
\label{moyal} \ee
is the Moyal bidifferential $\star$-operator.\\

To complete this brief summary of the Weyl-Wigner-Groenewold-Moyal formalism
note that if $A({\boldsymbol \Pi}, {\bf Q}, t)$ is a Heisenberg operator then
\be
A^{W} ({\bf p},{\bf q}, t) = \exp \{ - \frac{2t}{\hbar} H^{W}
\sin (\frac{\hbar}{2} \Lambda) \}  A^{W} ({\bf p},{\bf q}, 0), \label{heis}
\ee
so setting $A^{W} ({\bf p},{\bf q}, 0)$ equal to ${\bf p}$ and ${\bf q}$ we get
\ba
{\dot {\bf p}}&=& {\dot {\bf p}}^{W} (0) = H^{W} \Lambda {\bf p}^{W} (0) = -\nabla_{\bf q} H^{W}, \\
{\dot {\bf q}}&=& {\dot {\bf q}}^{W} (0) = H^{W} \Lambda {\bf q}^{W} (0) = \nabla_{\bf p} H^{W}, \label{hamilton}
\ea
respectively. Thus the c-numbers ${\bf p}$ and ${\bf q}$ satisfy Hamilton's
equations of motion, and may be interpreted as classical dynamical variables.\\
Note also, as it may be readily seen from (\ref{dist}), that the Wigner
distribution function is everywhere real and its projection on configuration and
momentum space gives the correct quantum mechanical configuration and
momentum probabilities, respectively. Hence its designation as a quasi-probability
density function.\\

Parallel to the classical phase-space integral equation
(\ref{wigner}), for the case when $|\Psi\rangle$
is a pure energy state there is a stronger equation, known as the
$\star$-value equation which can be derived directly from the
energy eigenvalue equation
\be
H({\bf \Pi}, {\bf Q})|\Psi\rangle= E |\Psi\rangle.\label{pure}
\ee
Indeed, using the fact that the $c$-Weyl function of a product of two
operators is equal to the Moyal product of their corresponding
$c$-Weyl functions ({\it cf.} (\ref{wigner})), we have that
\be
(H({\bf \Pi}, {\bf Q}){\boldsymbol \rho})^W = H^{W} \star \rho^W \label{pure2}
\ee
where $\rho^W$ on the right side of (\ref{pure2}) stands for the
$c$-Weyl equivalent to the density matrix ${\boldsymbol \rho}= |\Psi\rangle \langle\Psi|$.
Now, by (\ref{alpha}) and (\ref{weyl}), we can write
\be
H({\bf \Pi}, {\bf Q}){\boldsymbol \rho} =\int \alpha({\bf x},
{\bf y})e^{\frac{i}{\hbar}({\bf x}\cdot{\boldsymbol \Pi}+ {\bf y}\cdot {\bf Q})}d{\bf x} \ d{\bf y}
\ee
from where we derive
\be
\begin{split}
\alpha = (2\pi\hbar)^{-d}\langle\Psi|e^{-\frac{i}{\hbar}({\bf x}
\cdot{\boldsymbol \Pi}+ {\bf y}\cdot {\bf Q})}H |\Psi\rangle\\
=(2\pi\hbar)^{-d} E \int d{\bf q}^{\prime} \psi^{\ast}({\bf q}^{\prime})
\ e^{-\frac{i}{\hbar}{\bf y}\cdot({\bf q}^{\prime}-\frac{{\bf x}}{2})}
\psi({\bf q}^{\prime} -{\bf x}),\label{pure3}
\end{split}
\ee
and
\be
\begin{split}
(H {\boldsymbol \rho})^{W}=E (2\pi\hbar)^{-d}\int \int\int d{\bf x} \ d{\bf y}
 \ d{\bf q}^{\prime} e^{\frac{i}{\hbar}({\bf x}\cdot {\bf p}+{\bf y}\cdot {\bf q})}\\
\times \psi^{\ast}({\bf q}^{\prime})
\ e^{-\frac{i}{\hbar}{\bf y}\cdot({\bf q}^{\prime}-\frac{{\bf x}}{2})}
\psi({\bf q}^{\prime} -{\bf x}).\label{pure4}
\end{split}
\ee Integrating over ${\bf y}$ and ${\bf q}^{\prime}$ and
comparing with (\ref{dist}) we see that $\rho_W$ is precisely the
$c$-Weyl function corresponding to ${\boldsymbol \rho}$, so it
immediately follows that \be H^{W} ({\bf p},{\bf q})\star \rho_{W}
({\bf p},{\bf q}) = E \rho_{W} ({\bf p},{\bf q}).\label{starv} \ee
We emphasize here that $H^{W} ({\bf p},{\bf q})$ is in general not
equal to the $c$-function obtained by replacing the momentum and
position operators in the original quantum Hamiltonian by their
corresponding classical dynamical variables. This will be only
true for Hamiltonians of the form ${\boldsymbol \Pi}^{2}/2m +
V(\bf Q)$, and
will be an important {\it proviso} in our subsequent discussions.\\
Note also that by making use of the integral representation
\be
\begin{split}
    A^{W}_{1} ({\bf p}, {\bf q}) \star A^{W}_{2} ({\bf p}, {\bf q})  =
    (2\pi\hbar)^{-2d}\int\dots\int d{\bf p}^{\prime} \ d{\bf p}^{\prime\prime}
      d{\bf q}^{\prime} \ d{\bf q}^{\prime\prime}  A^{W}_{1} ({\bf p}^{\prime}, {\bf q}^{\prime})
       A^{W}_{2} ({\bf p}^{\prime\prime}, {\bf q}^{\prime\prime})\\
       \exp[-\frac{2i}{\hbar}({\bf p}\cdot({\bf q}^{\prime} - {\bf q}^{\prime\prime})
       + {\bf p}^{\prime}\cdot({\bf q}^{\prime\prime} - {\bf q}) +
        {\bf p}^{\prime\prime}\cdot({\bf q} - {\bf q}^{\prime}))], \label{int1}\\
\end{split}
\ee
it immediately follows that
\be
\int\int d{\bf p} \ d{\bf q} \ H^{W} ({\bf p}, {\bf q}) \star\rho_{W} ({\bf p}, {\bf q})
= \int\int d{\bf p} \ d{\bf q} \ H^{W} ({\bf p}, {\bf q})
 \rho_{W} ({\bf p}, {\bf q}) = E, \label{int2}
\ee
which is consistent with (\ref{wigner}).\\

Let us now turn to the noncommutative Heisenberg algebra
\begin{align}
[R_i, R_j]  &= i\hbar \theta_{ij}, \label{noncomm1} \\
[P_i, P_j] &= i\hbar {\bar \theta}_{ij},\label{noncomm2} \\
[R_i,P_j] &= i \hbar \delta_{ij}, \label{noncomm3}
\end{align} where
$\theta_{ij}$ and ${\bar \theta}_{ij}$ are evidently antisymmetric
matrices reflecting the noncommutativity of phase space. In order
to study the quantum mechanics associated with operators which are
arbitrary functions of ${\bf R}$ and ${\bf P}$, and in particular
their eigenvalues and eigenstates in the context of the
Weyl-Wigner-Groenewold-Moyal phase space formalism, we need first
to apply the quantum mechanical equivalent of the Seiberg-Witten
map to express the algebra of operators (\ref{noncomm1},
\ref{noncomm2}, \ref{noncomm3}) in terms of their ``commutative"
counterparts (\ref{heisen}). To this end, and making use of the
results in \cite{sochichiu} ({\it cf.} also \cite{smailagic},
\cite{smai-spallucci}, \cite{li}), we write a linear
representation of the algebra (\ref{noncomm1}), (\ref{noncomm2}),
(\ref{noncomm3}) as \be Q_i = a_{ij}R_j + b_{ij}\Pi_j
,\nonumber\ee \be P_i = c_{ij}R_j + d_{ij} \Pi_j  .
\label{mappings}\ee Substituting this expressions into
(\ref{noncomm1}), (\ref{noncomm2}), (\ref{noncomm3}) and using
(\ref{heisen}) one obtains the matrix equations \ba
AB^{T} - BA^{T} &=&  \Theta \nonumber\\
CD^T -DC^T &=& {\bar \Theta} \nonumber\\
AD^T -BC^T &=& {\bf 1},\label{matrix}
\ea
where the notation is self-evident. The solution of the above conditions
determine the structure of the mapping (\ref{mappings}).
For our present purposes we shall not be concerned with the problem of
finding and classifying general solutions to this problem.
It will suffice to consider one of the possible solutions which can
be readily found by choosing
$A=\lambda {\bf 1}, D=\mu {\bf 1}$, and also assuming that $B$ and $C$
are antisymmetric matrices. It is then easy to show
that
\be
B= -\frac{1}{2\lambda} \Theta,
\ee
and
\be
C = \frac{1}{2\mu} {\bar \Theta},
\ee
subject to the constraint
\be
{\bar \Theta} \Theta =\Theta{\bar \Theta} = 4\lambda\mu (\lambda\mu -1){\bf 1}.\label{cond1}
\ee
Thus we write
\ba
R_i &=& \lambda Q_i -\frac{1}{2\lambda} \theta_{ij}\Pi_j ,\label{sw1}\\
P_i &=& \mu\Pi_i + \frac{1}{2\mu} {\bar \theta}_{ij} Q_j ,\label{sw2}
\ea
where $\lambda$ and $\mu$ are constants.
Note that if we require
${\bf R}$ and ${\bf P}$ to be Hermitian, then $\lambda$, $\mu$,
$\theta_{ij}$ and ${\bar \theta}_{ij}$ have to be real.\\

Let us now investigate the implications of this specific noncommutative phase-space quantization
scheme by considering two exactly soluble problems.

\section{The Landau problem in noncommutative phase-space}

Neglecting spin, consider the 2-dimensional noncommutative phase-space
quantum Hamiltonian for an electron moving in a magnetic field $B$
in the direction normal to the quantum plane  $(R_1 , R_2)$:
\be
H({\bf P}, {\bf R})= \frac{1}{2m} ({\bf P} +\frac{e}{c} {\bf A})^2 .\label{landau}
\ee
In the symmetric gauge
\be
{\bf A} = (-\frac{B}{2} R_2 , \frac{B}{2} R_1 ), \label{gauge}
\ee
equation (\ref{landau}) reads, after substituting (\ref{sw1}), (\ref{sw2}),
\be
\begin{split}
H({\bf P}, {\bf R})= {\hat H}({\boldsymbol \Pi}, {\bf Q})= &
\frac{1}{2m} [(\mu -\frac{eB\theta}{4c\lambda})\Pi_1 +
(\frac{{\bar \theta}}{2\mu} -
 \frac{eB\lambda}{2c}) Q_2 ]^2 + \\
 &\frac{1}{2m} [(\mu - \frac{eB\theta}{4c\lambda})\Pi_2 - (\frac{{\bar \theta}}{2\mu} -
 \frac{eB\lambda}{2c}) Q_1 ]^2 ,\label{nonlandau}
 \end{split}
\ee
where we have also used $\theta_{ij}=\epsilon_{ij} \theta$ and
${\bar \theta}_{ij}=\epsilon_{ij} {\bar \theta}$.\\

Note now that by virtue (\ref{wigner}) the Weyl function associated
with the Hamiltonian (\ref{nonlandau}) is
\be
H^{W}({\bf p}, {\bf q})=  \frac{1}{2m} [(\mu + \frac{\kappa}{\lambda}) p_1 +
(\frac{{\bar \theta}}{2\mu} -
 \frac{eB\lambda}{2c}) q_2 ]^2 +  \frac{1}{2m} [(\mu + \frac{\kappa}{\lambda})p_2 -
 (\frac{{\bar \theta}}{2\mu} -
 \frac{eB\lambda}{2c}) q_1 ]^2 , \label{weyllandau1}
\ee
where
\be
\kappa:= -\frac{eB\theta}{4c}.\label{kapa}\\
\ee

We can now use this expression together with (\ref{starv}) to solve the
$\star$-value equation for the Wigner distribution function. We thus
have the second order differential equation
\be
\begin{split}
H^W \star \rho_W =& \left\{\frac{1}{2m} [(\mu +
\frac{\kappa}{\lambda}) (p_1 -\frac{i\hbar}{2}\partial_{q_1}) +
(\frac{{\bar \theta}}{2\mu} -
 \frac{eB\lambda}{2c}) (q_2 +\frac{i\hbar}{2}\partial_{p_2}) ]^2 + \right. \\
 & \left.\frac{1}{2m} [(\mu + \frac{\kappa}{\lambda})(p_2
-\frac{i\hbar}{2}\partial_{q_2}) -(\frac{{\bar \theta}}{2\mu} -
 \frac{eB\lambda}{2c}) (q_1 +\frac{i\hbar}{2}\partial_{p_1})]^2 \right\}\rho_W .
 \label{weyllandau2}
\end{split}
\ee Separating the real and imaginary parts in the above
expression, results in \be
\begin{split}
&\left\{\frac{1}{2m} [(\mu + \frac{\kappa}{\lambda}) p_1 +
(\frac{{\bar \theta}}{2\mu} - \frac{eB\lambda}{2c}) q_2 ]^2  +
\frac{1}{2m} [(\mu + \frac{\kappa}{\lambda})p_2 -(\frac{{\bar
\theta}}{2\mu} - \frac{eB\lambda}{2c}) q_1 ]^2 - \right. \\
&\frac{\hbar^2 }{8 m}[(\mu + \frac{\kappa}{\lambda})^2
\nabla_{{\bf q}} \cdot \nabla_{{\bf q}}
 + (\frac{{\bar \theta}}{2\mu} -  \frac{eB\lambda}{2c})^2 \nabla_{{\bf p}} \cdot
 \nabla_{{\bf p}} + \\& \left.
2(\mu + \frac{\kappa}{\lambda})(\frac{{\bar \theta}}{2\mu} -
\frac{eB\lambda}{2c}) (\partial_{p_1}\partial_{q_2}
-\partial_{p_2}\partial_{q_1})] \right\}\rho_W =E \rho_W ,
\label{real}
\end{split}
\ee

\be \label{im}
\begin{split}
-\frac{i\hbar}{2 m}[(\mu + \frac{\kappa}{\lambda})^2 {\bf p}\cdot
{\nabla_{{\bf q}}}+ (\mu + \frac{\kappa}{\lambda})(\frac{{\bar
\theta}}{2\mu} - \frac{eB\lambda}{2c})
 (q_2 \partial_{q_1} - q_1 \partial_{q_2}+ p_2 \partial_{p_1} - p_1
 \partial_{p_2})\\
 -(\frac{{\bar \theta}}{2\mu} -
 \frac{eB\lambda}{2c})^2 ({\bf q} \cdot \nabla_{\bf p})] \rho_W = 0.
\end{split}
\ee Now, since the time evolution of the Wigner function is given
by \be \frac{\partial \rho_{W}}{\partial t} = \frac{2}{\hbar}
H^{W} \sin \ (\frac{\hbar \Lambda}{2})\ \rho_W, \label{evol} \ee
and, since for a stationary system the density matrix ${\boldsymbol  \rho} =
|\Psi \rangle \langle \Psi|$ commutes with the Hamiltonian ${\hat
H}({\boldsymbol \Pi}, {\bf Q})$, it clearly follows that the right
side of (\ref{evol}) has to be zero. Furthermore, since the Weyl
function $H^{W}$ for the Landau Hamiltonian is at most quadratic
in the classical dynamical variables ({\it cf.}(\ref{weyllandau1}))
only the first term in the series expansion of the operator $\sin
\ (\frac{\hbar \Lambda}{2})$ contributes to (\ref{evol}). Hence
\be H^{W}  \ (\frac{\hbar \Lambda}{2})\ \rho_W =0. \label{ident}
\ee But this is precisely equation (\ref{im}). Noting, in
addition, that (\ref{ident}) would be identically satisfied if we
require that $\rho_W$ be a function of $H^{W}$, we shall now make
this ansatz and use (\ref{real}) to evaluate  $\rho_W$. By a
rather direct, albeit tedious calculation, we arrive at \be
-\frac{\hbar^2}{ m^2}(\mu + \frac{\kappa}{\lambda})^2 \
(\frac{{\bar \theta}}{2\mu} -  \frac{eB\lambda}{2c})^2 \
(\xi\frac{\partial^2 \rho_{W}} {\partial \xi^2} + \frac{\partial
\rho_{W}}{\partial \xi} ) +\xi\rho_W = E \rho_{W}, \label{realbis}
\ee
where we have set $\xi:= H^W$.\\
Moreover, letting
\be
\tau:= \frac{\hbar}{2m} (\mu + \frac{\kappa}{\lambda}) \ (\frac{{\bar \theta}}{2\mu} -
 \frac{eB\lambda}{2c}),\label{tau}
\ee
and introducing the new variable $\eta:=\frac{\xi}{\tau}$ we get, from (\ref{realbis}),:
\be
\eta \frac{\partial^2 \rho_{W}}{\partial \eta^2} +
\frac{\partial \rho_{W}}{\partial \eta} - (\frac{\eta}{4} -
\frac{E}{4\tau})\ \rho_W =0.\label{real2}
\ee

Making the additional change of dependent variable \be \rho_W =
e^{-\frac{\eta}{2}} \ \omega, \label{depv} \ee equation
(\ref{real2}) takes the form of Laguerre's differential equation
\be [\eta \frac{\partial^2 }{\partial \eta^2} + (1 - \eta)
\frac{\partial }{\partial \eta} + \frac{E}{4\tau} -\frac{1}{2}]\
\omega =0, \label{laguerre} \ee which, for integral values of
$\frac{E}{4\tau} -\frac{1}{2} = n$, has a solution in terms of
Laguerre polynomials \be \omega= L_{n}(\eta)=\sum_{k=0}^{n} (-1)^k
\left( \begin{array}{c} n\\k \end{array} \right)
\frac{\eta^k}{k!}.\label{plynom} \ee The energy spectrum for the
Landau problem is then given by \be E= \frac{2\hbar}{m} (\mu +
\frac{\kappa}{\lambda}) \ (\frac{{\bar \theta}}{2\mu} -
\frac{eB\lambda}{2c}) (n+\frac{1}{2}), \label{energy} \ee and the
corresponding Wigner distribution function by \be \rho_W =
\exp{(-\frac{H^W}{2\tau})} \
L_{n}(\frac{H^W}{\tau}),\label{wignerf} \ee
with $\tau (\theta, {\bar \theta}, \lambda, \mu)$ given by (\ref{tau}).\\

Let us now compare the above results with others appearing in the
literature for the Landau and similar problems obtained by
applying a certain deformation quantization prescription to the
point product of a classical Hamiltonian and the Wigner function.
To be more specific, in the Landau problem for example (see {\it
e.g.} \cite{dayi}), the classical Hamiltonian is taken to be the
one determined by (\ref{landau}) and (\ref{gauge}) with the
operators ${\bf R}$ and ${\bf P}$ replaced by the classical
phase-space variables and, in order to take care of the
noncommutativity of the phase-space, the $\star$-value equation
(\ref{starv}) is replaced by the prescription \be H({\bf p}, {\bf
q}) \star^{\prime} \rho_W = E \rho_W, \label{qstarv} \ee where \be
\star^{\prime} \equiv \star_{\hbar} \star_{\theta} \star_{{\bar
\theta}}, \label{def} \ee \be \star_{\theta} = \exp [
\frac{i\hbar}{2}\sum_{i,j}\theta_{ij} (
{\overleftarrow{\partial_{q_i}}}\cdot
{\overrightarrow{\partial_{q_j}}} -
{\overleftarrow{\partial_{q_j}}}\cdot
{\overrightarrow{\partial_{q_i}}} ) ],
 \label{starq}
\ee  \be \star_{\bar{\theta}} = \exp [
\frac{i\hbar}{2}\sum_{i,j}
{\bar\theta}_{ij}({\overleftarrow{\partial_{p_i}}}\cdot
 {\overrightarrow{\partial_{p_j}}}
- {\overleftarrow{\partial_{p_j}}}\cdot
{\overrightarrow{\partial_{p_i}}} ) ], \label{starp} \ee and
$\star_{\hbar}$ is the Moyal $\star$-operator defined in
(\ref{moyal}). Note that this approach hinges on the criterion
that the noncommutative algebra  (\ref{noncomm1}, \ref{noncomm2},
\ref{noncomm3}) can be derived via the composition
$\star_{\theta} \star_{\bar \theta}$ in (\ref{def}).\\

For the particular case when ${\bar \theta}=0$ ($[P_i, P_j]=0$),
the energy eigenstates and Wigner function for the 2-dimensional
Landau problem obtained with the prescription (\ref{qstarv}) and
those obtained with the formalism described before (Eqs. (\ref{energy})
and (\ref{wignerf})) turn out to be the same. The
reason becomes obvious when we note that when acting with the
operator $\star_{\theta}$ from the right on the classical
Hamiltonian yields the operator \be {\hat H}_{nc} = \frac{1}{2m}[
( (1+\kappa){\hat p_1} - \frac{eB}{2c}{\hat q}_2 )^2 + (
(1+\kappa){\hat p_2} + \frac{eB}{2c}{\hat q}_1 )^2 ],\label{ef1}
\ee where ${\hat p_1}$, ${\hat p_2}$ and ${\hat q_1}$, ${\hat
q_2}$ are momenta and position operators, respectively, in the
coordinate representation. Defining an effective $c$-Hamiltonian
by replacing these operators by their corresponding $c$-dynamical
variables, results in the effective $c$-number Hamiltonian \be
H_{eff} = \frac{(1+\kappa)^2}{2m}{\bf p}^2 + \frac{m \omega^2}{8}
{\bf q}^2 + \frac{(1+\kappa)\omega}{2} (q_{1} p_2 - q_{2} p_1
),\label{ef2} \ee with $\omega = \frac{eB}{mc}$. \ But (\ref{ef2})
is for this particular case the same as the Weyl function that we
would get from the Weyl-Wigner-Groenewold-Moyal formalism. Indeed,
by virtue of the condition (\ref{cond1}), the constants $\lambda$
and $\mu$, appearing as a result of the transformations
(\ref{sw1}) and (\ref{sw2}), are related by \be \mu = \frac{1\pm
\sqrt{1-\theta {\bar \theta}}}{2\lambda}. \label{cond2} \ee So
that when ${\bar \theta}=0$, $\mu$ and $\lambda$ both need
necessarily be equal to $1$, and the Weyl function $H^W$ derived
in (\ref{weyllandau1}) turns out to be the same (after setting
$\mu=1$, $\lambda=1$, and ${\bar \theta}=0$) to the effective
Hamiltonian (\ref{ef2}). This is of course not true for the more
general cases where the Weyl equivalent to a
quantum operator is different from the classical operator.\\
Furthermore, when $\theta,{\bar \theta}\neq 0$ it also follows clearly
from (\ref{cond2}) that $\lambda$ and $\mu$ can not be chosen
simultaneously to be equal to $1$. Hence the results obtained for
the energy eigenvalues and the Wigner function will be
quantitatively quite different for the two approaches (compare with
results in \cite{dayi}), and in fact
the correct procedure is the one which uses the mappings
(\ref{sw1}), (\ref{sw2}) and the $\star$-value
equation (\ref{starv}) leading to equations
(\ref{energy}) and (\ref{wignerf}).\\

\section{The harmonic oscillator in noncommutative phase-space}

Another quantum mechanical problem on the noncommutative plane
that has been extensively considered in the literature is that of
a particle in an external central potential described by the
Hamiltonian \be H({\bf P}, {\bf R}) = \frac{{\bf P}^2}{2m} +
V(|{\bf R}|^2), \label{pot} \ee where ${\bf P}$ and ${\bf R}$
satisfy the algebra (\ref{noncomm1}, \ref{noncomm2},
\ref{noncomm3}). Note in particular that for a free particle the
mapping (\ref{sw2}) leads back to the Landau problem considered in
the previous section when we identify ${\bar \theta}$ with the
external constant
magnetic field. \\

>From the extended noncommutative phase-space point of view of the
Weyl-Wigner-Groenewold-Moyal formalism, general solutions to
(\ref{pot}) for the energy spectrum and Wigner functions can
become quite complicated depending on the form of the potential.
One reason for this is that even when $V(|{\bf R}|^2)= V((\lambda
Q_i - \frac{1}{2\lambda} \theta_{ij} \Pi_j)^2)$ is of a polynomial
form in the argument, it clearly follows that $$(|{\bf R}|^{2m +
2n})^W = (|{\bf R}|^{2m})^W \star (|{\bf R}|^{2n})^W  \neq (|{\bf
R}|^{2m})^W  (|{\bf R}|^{2n})^W $$ except for the case when $m=n$.
Hence the Weyl $c$-functions corresponding to the potential part
of the Hamiltonian are not, in general, just the classical
functions resulting from replacing the operators ${\bf Q}$ and
${\boldsymbol \Pi}$ by their corresponding classical canonical
variables. This will only be so for polynomial functions of the
form $V(|{\bf R}|^2)= \sum_{n} a_n |{\bf R}|^{2^n}$. It is not our
objective here however to pursue the discussion for the general
case, as it will suffice for our purposes to concentrate on the
problem of the harmonic oscillator in the noncommutative
phase-space plane. We shall therefore consider the quantum
Hamiltonian \be
\begin{split}
H({\bf P}, {\bf R}) = \frac{{\bf P}^2}{2m} +\frac{m\omega^2}{2}|{\bf R}|^2 = \frac{1}{2m}(\mu\Pi_1
+\frac{1}{2\mu}{\bar \theta} Q_2)^2 + \frac{1}{2m}(\mu\Pi_2
-\frac{1}{2\mu}{\bar \theta} Q_1)^2 \\
 + \frac{m\omega^2}{2} (\lambda Q_1 -\frac{1}{2\lambda}\theta\Pi_2)^2 + \frac{m\omega^2}{2}
 (\lambda Q_2 +\frac{1}{2\lambda}\theta\Pi_1)^2.
\label{ho}
\end{split}
\ee
Rearranging terms, (\ref{ho}) reads
\be
H({\bf P}, {\bf R}) = \alpha^2 {\bf Q}^2 + \beta^2 {\boldsymbol \Pi}^2 +
(\frac{{\bar \theta}}{2m} +\frac{m\omega^2 \theta}{2})(\Pi_1 Q_2 - \Pi_2 Q_1),
\label{ho1}
\ee
where
\be
\alpha^2 = (\frac{\lambda^2 m\omega^2}{2} + \frac{\bar \theta^2}{8m\mu^2}),\label{alfa}
\ee
\be
\beta^2 = (\frac{\mu^2}{2m} + \frac{m\omega^2 \theta^2}{8\lambda^2}). \label{beta}
\ee
Introducing now the creation and annihilation operators
\ba
{\hat a_i}^{\dagger}&=& \frac{\alpha}{\sqrt {2\hbar\alpha\beta}} Q_i -
i\frac{\beta}{\sqrt {2\hbar\alpha\beta}}\Pi_i ,\\
{\hat a_i} &=& \frac{\alpha}{\sqrt {2\hbar\alpha\beta}} Q_i +
i\frac{\beta}{\sqrt {2\hbar\alpha\beta}}\Pi_i ,
\ea
we can write (\ref{ho1}) as
\be
\begin{split}
H({\bf P}, {\bf R}) = 2\hbar\alpha\beta ({\hat a_1}^{\dagger} {\hat a_1} +
{\hat a_2}^{\dagger} {\hat a_2} +1) \\
-i\hbar(\frac{{\bar \theta}}{2m} + \frac{m\omega^2}{2} \theta)
({\hat a_1} {\hat a_2}^{\dagger}  -{\hat a_1}^{\dagger}
{\hat a_2} ).\label{ho2}
\end{split}
\ee
Note that in the above
\be
[{\hat a_i},  {\hat a_j}^{\dagger}]=\delta_{ij}, \label{comeq1}
\ee
\be
[{\hat a_i},  {\hat a_j}]=[{\hat a_i}^{\dagger}, {\hat a_j}^{\dagger}]=0. \label{comeq2}
\ee
Note also that the angular momentum term
\be
L= ({\hat a_1} {\hat a_2}^{\dagger}  -{\hat a_1}^{\dagger}{\hat a_2} ) \label{angm}
\ee
in (\ref{ho2}) commutes with the number operator $ N ={\hat a_1}^{\dagger} {\hat a_1} +
{\hat a_2}^{\dagger} {\hat a_2} $
and so it is a constant of the motion, and they both together form a complete
set of commuting observables. Indeed, introducing the
new annihilation and creation operators \cite{messiah}
\ba
{\hat A}_{\pm}&=&\frac{1}{\sqrt 2}({\hat a_1} \mp i{\hat a_2}),\\
{\hat A}^{\dagger}_{\pm}&=&\frac{1}{\sqrt 2}({\hat a_1}^{\dagger} \pm i{\hat a_2}^{\dagger} ),
\label{newop}
\ea
which satisfy the commutation relations
\be
[{\hat A}_{\pm}, {\hat A}_{\pm}]= [{\hat A}_{\pm}, {\hat A}_{\mp}]=0,
\ee
\be
[{\hat A}_{\pm}^{\dagger}, {\hat A}_{\pm}^{\dagger}]= [{\hat A}_{\pm}^{\dagger},
{\hat A}_{\mp}^{\dagger}]=0,
\ee
\be
[{\hat A}_{\pm}, {\hat A}_{\pm}^{\dagger}]=1,\ [{\hat A}_{\mp}, {\hat A}_{\mp}^{\dagger}]=1,
\ee
\be
[{\hat A}_{\mp}, {\hat A}_{\pm}^{\dagger}]=0, \ [{\hat A}_{\pm}, {\hat A}_{\mp}^{\dagger}]=0.
\ee
we have that the number operators
\ba
N_{+} &=& {\hat A}^{\dagger}_{+} {\hat A}_{+}\\
N_{-} &=& {\hat A}^{\dagger}_{-} {\hat A}_{-},\label{ns}
\ea
form a complete set of commuting observables, whose spectra is the sequence of non-negative integers
$$n_{+}=0,1,\dots  \;\;\; \ n_{-}=0,1,\dots ,$$
respectively. Their common eigenstates are
\be
|n_{+}n_{-}\rangle =( n_{+}! n_{-}!)^{-\frac{1}{2}} ({\hat A}^{\dagger}_{+})^{n_{+}}
({\hat A}^{\dagger}_{-})^{n_{-}} |00\rangle , \label{eigenstates}
\ee
such that
\be
N_{\pm} |n_{+}n_{-}\rangle= n_{\pm} \ |n_{+}n_{-}\rangle. \label{neigen}
\ee
We can therefore write

\be
H({\bf P}, {\bf R}) = 2\hbar\alpha\beta (N_{+} + N_{-} +1) -\hbar (\frac{{\bar \theta}}{2m} +
\frac{m\omega^2}{2} \theta) (N_{+} - N_{-}),\label{ho7}
\ee
and
\be
\begin{split}
H({\bf P}, {\bf R}) |n_{+}n_{-}\rangle =\hspace{2.5in}\\
[ 2\hbar\alpha\beta (n_{+} + n_{-} +1) -\hbar (\frac{{\bar \theta}}{2m} +
\frac{m\omega^2}{2} \theta) (n_{+} - n_{-})] \ |n_{+}n_{-}\rangle.\label{ho8}
\end{split}
\ee Let us now denote by ${\bar A}_{\pm}$ and $A_{\pm}$ the
classical Weyl-equivalents to the operators  ${\hat
A}^{\dagger}_{\pm}$, ${\hat A}_{\pm}$, respectively. In this
holomorphic coordinates the Moyal $\star$-operator is given by \be
\star = \exp[\frac{1}{2}({\overleftarrow \partial}_{A_{+}}
{\overrightarrow \partial}_{{\bar A}_{+}}-{\overleftarrow
\partial}_{ {\bar A}_{+}} {\overrightarrow \partial}_{A_{+}} +
{\overleftarrow \partial}_{A_{-}}{\overrightarrow \partial}_{{\bar
A}_{-}}-{\overleftarrow \partial}_{ {\bar A}_{-}} {\overrightarrow
\partial}_{A_{-}} )].\label{holomoyal2} \ee We thus have that \be
{\bar A}_{\pm} \star A_{\pm} = {\bar A}_{\pm}  A_{\pm}
-\frac{1}{2}, \ee and the Weyl $c$-function corresponding to
(\ref{ho7}) is \be H^W (\xi_1 ,\xi_2) = 2\hbar\alpha\beta (\xi_1 +
\xi_2) -\hbar (\frac{{\bar \theta}}{2m} + \frac{m\omega^2}{2}
\theta) (\xi_1 - \xi_2),\label{ho9} \ee
where $\xi_1 := {\bar A}_{+}  A_{+}$ and  $\xi_2 := {\bar A}_{-}  A_{-}$.\\
Setting
\be
\gamma :=(\frac{{\bar \theta}}{2m} + \frac{m\omega^2}{2} \theta), \label{gama}
\ee
and rearranging terms, we can write
\be
H^W (\xi_1 ,\xi_2) =(2\hbar\alpha\beta -\hbar\gamma)\xi_1 +
(2\hbar\alpha\beta +\hbar\gamma)\xi_2 .\label{ho10}
\ee
It is easy to see that for this particular form of the Weyl-Hamiltonian
function the $\star$-value equation (\ref{starv})
yields
\be
\begin{split}
H^W (\xi_1 ,\xi_2) \star \rho_W = [(2\hbar\alpha\beta -\hbar\gamma)\xi_1 + (2\hbar\alpha\beta +\hbar\gamma)\xi_2 +\\
\frac{1}{2}(2\hbar\alpha\beta -\hbar\gamma)({\bar A}_{+}\frac{\partial}{\partial {\bar A}_{+}} - { A}_{+}\frac{\partial}{\partial { A}_{+}})+\\
\frac{1}{2}(2\hbar\alpha\beta + \hbar\gamma)({\bar A}_{-}\frac{\partial}{\partial {\bar A}_{-}} - { A}_{-}\frac{\partial}{\partial { A}_{-}})+\\
-\frac{1}{4}(2\hbar\alpha\beta -\hbar\gamma)\frac{\partial^2}{\partial A_{+}\partial {\bar A}_{+}}
-\frac{1}{4}(2\hbar\alpha\beta +\hbar\gamma)\frac{\partial^2}{\partial A_{-} \partial {\bar A}_{-}}\\
-E]\rho_W =0.\label{starv4}
\end{split}
\ee
The above equation can now be readily solved for the energy spectrum and Wigner function by
separation of variables and by following a
procedure similar to that used in Sec.3. We thus get the set of ordinary differential equations
\be
[\xi_1 -\frac{1}{4}(\frac{\partial}{\partial \xi_1} +
\xi_1 \frac{\partial^2}{{\partial \xi_1}^2})-\varepsilon_1 ]U(\xi_1)=0 ,\label{sol1}
\ee
\be
[\xi_2 -\frac{1}{4}(\frac{\partial}{\partial \xi_2} +
\xi_2 \frac{\partial^2}{{\partial \xi_2}^2})-\varepsilon_2 ]V(\xi_2)=0, \label{sol2}
\ee
where $\rho_W = U(\xi_1) V(\xi_2)$, and $(2\hbar\alpha\beta -
\hbar\gamma)\varepsilon_1 + (2\hbar\alpha\beta
+\hbar\gamma)\varepsilon_2 = E$.\\
The explicit solutions to (\ref{sol1}) and (\ref{sol2}) in terms of Laguerre polynomials are
\be
U(\xi_1) = e^{-2\xi_1} \ L_{n_1} (4\xi_1), \label{partial1}
\ee

\be
V(\xi_2) = e^{-2\xi_2} \ L_{n_2} (4\xi_2), \label{partial2}
\ee
where $n_1, n_2$ are non-negative integers, and
\be
\varepsilon_1 = (n_1 +\frac{1}{2}), \;\;\;\;\;  \varepsilon_2 = (n_2 +\frac{1}{2}).
\ee
Hence
\be
E= (2\hbar\alpha\beta)(n_1+n_2 +1) +\hbar\gamma (n_1 -n_2),\label{energyho}
\ee
and, in terms of canonical phase-space dynamical variables,
\be
\begin{split}
\rho_W = \exp[-(\frac{\alpha}{\beta}{\bf q}^2 +
\frac{\beta}{\alpha}{\bf p}^2 )] L_{n_1} (\frac{2}{\hbar}[\frac{\alpha}{\beta}{\bf q}^2
 + \frac{\beta}{\alpha}{\bf p}^2 + 2(q_1 p_2 -q_2 p_1)])\\
\times L_{n_2} (\frac{2}{\hbar}[\frac{\alpha}{\beta}{\bf q}^2
 + \frac{\beta}{\alpha}{\bf p}^2 - 2(q_1 p_2 -q_2 p_1)]). \label{wignerho}
\end{split}
\ee
Substituting (\ref{alfa}), (\ref{beta}) and (\ref{gama}) into (\ref{energyho})
we arrive at the final following expression for the
energy spectrum of the harmonic oscillator problem in noncommutative phase-space
\be
E= \pm\frac{\hbar}{2} [\sqrt {4\omega^2 + (m\omega^2 \theta -
\frac{{\bar \theta}}{m})^2} \ (n_1+n_2 +1) + (\frac{{\bar \theta}}{m}
+m\omega^2 \theta)(n_1-n_2)].\label{energyho2}
\ee
This expression is in agreement with that reported in the literature by
other authors ( see {\it e.g.} \cite{nair},
\cite{bellucci}) who derived it essentially by splitting the
algebra (\ref{noncomm1}), (\ref{noncomm2}), (\ref{noncomm3})
into two independent subalgebras and solving the quantum energy
eigenvalue equation after performing a Bogolyubov transformation.
There are however a few remarks
that should be made here. First, in the above mentioned papers the authors
consider three  possible cases which,
in our notation, correspond to $\kappa=1 - \theta{\bar \theta}=0$
(the so called ``critical point" case),  and $\kappa>0$, $\kappa<0$ .\\
The solution (\ref{energyho2}) corresponds to the case $\kappa>0$.
For the ``critical point" case ( $\kappa=1 - \theta{\bar
\theta}=0$) we obtain, as so do the authors in the above mentioned
references, that $\alpha\beta = \frac{1}{4}(m\omega^2 \theta
+\frac{1}{\theta m})$ so that the energy spectrum reduces to that
of a single harmonic oscillator: \be E = \hbar (m\omega^2 \theta
+\frac{1}{m\theta})(n_1 +\frac{1}{2}). \label{critical} \ee
Furthermore, the phase-space volume elements in the coordinates
${\bf R}, {\bf P}$ in the noncommutative Heinsenberg algebra are
related to the commutative ones ${\bf Q}, {\boldsymbol \Pi}$, by
\be dR_1 \ dR_2 \ dP_1 \ dP_2 = \left|J\left(\frac{{\bf R}, {\bf
P}}{{\bf Q}, {\boldsymbol \Pi}}\right)\right| \ dQ_1 \ dQ_2 \
d\Pi_1 \ d\Pi_2 , \ee and since by (\ref{sw1}), (\ref{sw2}), the
Jacobian turns out to be zero for this case it follows that the
density of states for a fixed energy becomes degenerate. Hence the
designation of ``critical point" for
this particular situation.\\
Note on the other hand that the additional
restriction (\ref{cond2}) implied by the mappings (\ref{sw1}), (\ref{sw2}) in our
formalism, precludes the case $\kappa<0$, since
the parameters $\lambda$ and $\mu$ are required by hermicity to be real.
Furthermore, while these parameters are irrelevant to
the energy spectrum problem, as they do not appear in the final expression,
this is clearly not so for the Wigner function (\ref{wignerho})
and the energy eigenstates for the problem. We thus have a complete
1-free-parameter set of solutions which lead to the same energy
spectrum for the harmonic oscillator problem but, by virtue of the
expectation value equation ({\it c.f.} (\ref{wigner}))
\be
\langle \Omega ({\bf P}, {\bf R}) \rangle =
\int\int d{\bf p} d{\bf q} \ \rho_W \ \Omega^W ({\bf p}, {\bf q}), \label{exp}
\ee
the spectrum of other observables of the theory may be dependent on this parameter. This, as well as
 its possible physical implications, remain
to be investigated. \\

In \cite{hatzi} the harmonic oscillator problem is considered from the point of
view of quantum deformation via the prescription
(\ref{def}). Here again, as in the case of the Landau problem in noncommutative
phase-space, discussed at the end of the previous
section, we are met both with the same conceptual and computational differences
in the derivation of the energy spectrum and the Wigner
function. First, of all the algebra $[R_1, R_2] = i\hbar\theta$, \ $[P_1, P_2] =- i\hbar\theta$, \ $[Q_i, P_j]=i\hbar\delta_{ij}$, initially
considered in that work is incompatible with the
mappings (\ref{sw1}), (\ref{sw2}) and the condition (\ref{cond2}). Second the
calculation of the Wigner function obtained for this
case as well as for the more general noncommutative phase-space algebra by means
of the Weyl-Moyal correspondence (\ref{qstarv})
leads to results quite different to (\ref{wignerho}), for the
same reasons as those discussed at the end of Sec.3. It is (\ref{wignerho})
that gives the correct quantum mechanics for the problem,
which again exemplifies our contention that it is the extended
Weyl-Wigner-Groenewold-Moyal formalism the correct procedure to follow when
considering these type of problems.

\section{Discussion and conclusions}

We have constructed a quantum mechanics over the noncommutative
phase-space $\{\bf R,P \!\}$, whose algebra is given by the
commutators in (\ref{noncomm1}), (\ref{noncomm2}),
(\ref{noncomm3}), by extending the Weyl-Wigner-Groenewold-Moyal
formalism with the
 mappings (\ref{sw1}) and (\ref{sw2}) which can be viewed as the quantum
 mechanical equivalent of the Seiberg-Witten map in
field theory. In this way operators defined over the quantum
variables $\{ {\bf R}, {\bf P}\! \}$ are first re-expressed in
terms of the ordinary quantum mechanics position and momentum
operators $\{ {\bf Q}, {\boldsymbol \Pi} \}$ and then their
corresponding $c$-Weyl equivalents are constructed by following
the usual procedures of
the Weyl-Wigner-Groenewold-Moyal formalism.\\
In particular, given a quantum Hamiltonian $H( {\bf P}, {\bf R}, t)$ which
determines the time evolution of the system,
the above procedure can be used to obtain its $c$-Weyl equivalent which in turn
can be used in the $\star$-value equation
(\ref{starv}) to derive the Wigner distribution function for the problem under
consideration. We stress here the fact,
as was elaborated in the text, that the $c$-Weyl equivalent of the original
Hamiltonian quantum operator is not in general equal to the $c$-function
resulting from replacing the
operators $ {\bf Q}, {\boldsymbol \Pi}$ in the former by their corresponding
classical canonical dynamical
variables.\\
We have applied the above considerations to two exactly soluble problems
and have specifically shown that
the use of the Weyl-Moyal equivalence, as given in (\ref{qstarv}), leads to
different results for the energy spectrum and
the Wigner function for these problems, thus verifying our contention that it
is either the intrinsically noncommutative
operator space approach or the extended Weyl-Wigner-Groenewold-Moyal
formalism the appropriate ones for constructing the quantum mechanics over the
noncommutative phase-space. Furthermore
since, as noted in the introduction, the former is hard to implement in  explicit
calculations for non-exactly soluble problems,
the study of the noncommutative
effects by means of perturbations can be best carried out via series expansions of
the Weyl and Wigner functions in the extended
 Weyl-Wigner-Groenewold-Moyal formalism.\\

 The essential difference between the approach advocated here of extending to
noncommutative phase-space the Weyl-Wigner-Groenewold-Moyal
formalism, and the
prescription for deformation quantization contained in equations
(\ref{qstarv}) and (\ref{def}) is that the former is unequivocal in
the sense that to a given quantum operator with arguments in the
algebra (\ref{noncomm1}, \ref{noncomm2}, \ref{noncomm3}) there
corresponds a unique Weyl function determined by (\ref{alpha}) and
(\ref{weyl}). For a given quantum Hamiltonian, it is this Weyl
function and the $\star$-value equation (\ref{starv}) that we
claim give
the correct Wigner function and
energy eigenvalues for the problem under consideration.\\

We have shown that when ${\bar \Theta}=0$ in the Heisenberg algebra
(\ref{noncomm1}, \ref{noncomm2}, \ref{noncomm3}),
there is at least one solution $(\lambda=\mu=1)$ of equations (\ref{matrix})
for which the Wey-Moyal correspondence (\ref{qstarv}), (\ref{def}), gives
quantizations equivalent to the extended Weyl-Wigner-Groenewold-Moyal
formalism, for the problems considered. We have also shown,
however, that this is a consequence of the particular situation stemming
from the fact that the $c$-Weyl functions related to the specific quantum Hamiltonians
are indeed those resulting from replacing the ${\boldsymbol \Pi}$ and ${\bf Q}$ operators
by their corresponding classical dynamical variables. In more general cases the two
quantization schemes would not be equivalent, even for the ${\bar \Theta}=0$ noncommutative
Heisenberg algebra. For the ${\bar \Theta}\neq 0$ case there seems not to be much sense in using
(\ref{qstarv}) to derive the Wigner function, since by this procedure the first two $\star$-products
in the composition (\ref{def}) lead to no effective classical
Hamiltonian in terms of canonical dynamical variables that would give sense to the third Moyal product
in the composition and hence to a phase-space quantum mechanics.\\

Another issue that was mentioned cursively in the text and needs further
investigation is the analysis and classification of the more
general solutions to the set of conditions (\ref{matrix}), and their possible
physical implications.\\

We conclude by remarking that deformation quantization would be of
course the natural procedure to follow when given a classical
Hamiltonian over classical phase-space one would try to infer the
corresponding noncommutative quantum one by some $\star$-operator.
In the context of deformation quantization one starts from a pair
of $c$-functions of the classical dynamical variables and quantum
deforms its point product by means of a $\star$-multidifferential
operator. There are many possible choices for these operators that
satisfy the usual properties of associativity, classical and
semi-classical limits. The universal one being the Kontsevich
product. Here one would also have to deal with the associated
operator ordering problems, in addition to the different possible
choices
of the $\star$-product.\\


\begin{thebibliography}{99}
\bibitem{rovelli}
Rovelli C and Smolin L., Editors 1995
\newblock Special Issue on Quantum Geometry and Diffeomorphism Invariant Quantum
Field Theory,
\newblock { J.\ Math.\ Phys.}  {\bf 36}.

\bibitem{connes}
Connes A 1994
\newblock Noncommutative Geometry,
\newblock {\em Academic  Press}
\newblock {San Diego, California}.

\bibitem{rivelles}
V.~O.~Rivelles,
Phys.\ Lett.\ B {\bf 558}, 191 (2003) [arXiv:hep-th/0212262].

\bibitem{alvarez}
L.~Alvarez-Gaume and S.~R.~Wadia,
Phys.\ Lett.\ B {\bf 501}, 319 (2001) [arXiv:hep-th/0006219].


\bibitem{seiberg}
N.~Seiberg and E.~Witten,
JHEP {\bf 9909}, 032 (1999) [arXiv:hep-th/9908142].



\bibitem{nair}
Nair V P and Polychronakos A P 2001
\newblock Quantum Mechanics on the Noncommutative Plane ad Sphere,
\newblock {\em  Phys.\ Lett.\ B}  {\bf 505}, 267-274
\newblock \texttt{hep-th/0011172}


\bibitem{sochichiu}
C.~Sochichiu,
arXiv:hep-th/0010149.

\bibitem{smailagic}
A.~Smailagic and E.~Spallucci,
Phys.\ Rev.\ D {\bf 65}, 107701 (2002) [arXiv:hep-th/0108216].

\bibitem{li}
K.~Li, J.~h.~Wang and C.~Chen,
arXiv:hep-th/0409234.

\bibitem{chaichian}
M.~Chaichian, M.~M.~Sheikh-Jabbari and A.~Tureanu,
Phys.\ Rev.\ Lett.\  {\bf 86}, 2716 (2001) [arXiv:hep-th/0010175].

M.~Chaichian, M.~M.~Sheikh-Jabbari and A.~Tureanu,
Eur.\ Phys.\ J.\ C {\bf 36}, 251 (2004) [arXiv:hep-th/0212259].

\bibitem{smai-spallucci}
A.~Smailagic and E.~Spallucci,
J.\ Phys.\ A {\bf 35}, L363 (2002) [arXiv:hep-th/0205242].

\bibitem{dayi}
O.~F.~Dayi and L.~T.~Kelleyane,
Mod.\ Phys.\ Lett.\ A {\bf 17}, 1937 (2002)
[arXiv:hep-th/0202062].

\bibitem{messiah}
A.~Messiah, Quantum Mechanics, Dover Publications 2000.

\bibitem{bellucci}
S.~Bellucci and A.~Nersessian,
Phys.\ Lett.\ B {\bf 542}, 295 (2002) [arXiv:hep-th/0205024].

\bibitem{hatzi}
A.~Hatzinikitas and I.~Smyrnakis,
J.\ Math.\ Phys.\  {\bf 43}, 113 (2002) [arXiv:hep-th/0103074].




\end{thebibliography}
\end{document}